\documentclass[a4paper]{jpconf-kb}
\usepackage{amsmath}
\usepackage{iopams}

\usepackage{graphicx}

\def\xrightarrow#1#2#3#4{\,\lower#1pt\hbox{$\stackrel{\stackrel{\displaystyle #2}%
{\hbox to #3cm{\rightarrowfill}}}{#4}$}\,}
        
\begin{document}

\begin{flushleft}
KCL-PH-TH/2013-25
\end{flushleft}

\title{Inflation and cosmic (super)strings: implications of their intimate relation revisited}

\author{M.\ Sakellariadou$^1$}

\address{$^1$ Department of Physics, King's College London, University of London,\newline
\phantom{$^1$ } Strand, London WC2R 2LS, United Kingdom}

\ead{mairi.sakellariadou@kcl.ac.uk}


\begin{abstract}
We briefly discuss constraints on supersymmetric hybrid inflation models and examine the consistency of brane inflation models. We then address
the implications for inflationary scenarios resulting from  the strong constraints  on the cosmic (super)string
tension imposed from the most recent cosmic microwave background temperature
anisotropies data.
\end{abstract}

\section{Introduction}
Cosmological inflation
\cite{Starobinsky:1980te,Guth:1980zm,Linde:1981mu}
offers,
by construction, a simple way to evade some of the problems related to
the initial conditions that plague the hot big bang model. In
addition, inflation 
can provide a simple
mechanism to get adiabatic perturbations that seem to lead to the
observed large scale structure and the measured Cosmic Microwave
Background (CMB)
temperature anisotropies. However, despite efforts over more than
three decades, inflation
remains a paradigm
in search of a model. Apart the issues related to the onset of
inflation
\cite{Calzetta:1992gv,Calzetta:1992bp,Germani:2007rt}, particle
physics models indicate difficulties in order to satisfy the required
properties of the scalar field that would play the r\^ole of the
inflaton.

The main bulk of efforts to realise an inflationary scenario is
concentrated on a scalar (single or multiple) field type of
inflation.
The simple minimal case of a
Higgs field driven inflation
\cite{Bezrukov:2007ep}
faces important difficulties
\cite{Buck:2010sv}
and inconsistencies
\cite{Barbon:2009ya, Atkins:2010yg}.
In the context of supersymmetric Grand
Unified Theories (GUTs),
 hybrid
inflationary models end with the formation of cosmic strings
\cite{Jeannerot:2003qv},
which to be compatible with the CMB
measurements require fine tuning of the free parameters
(couplings and mass scales) of the inflationary potential
\cite{Rocher:2004et,Rocher:2004my,Rocher:2006nh,Battye:2010hg}.
Moreover,
the singlets of the Standard Model (SM) seem to encounter difficulties
in order to play the r\^ole of the inflaton,
\%index[subject]{inflation},
as it has been recently shown
\cite{giaccomo-mairi}
in the case of the minimal SO(10) model.
 
One can argue that inflation
must be studied
in the context of an Ultra-Violet (UV) complete theory and as such,
string theory
is a well-studied
candidate. Within this framework, brane inflation
studied, lead to the formation of cosmic
superstrings
whose observational
consequences may offer a way of constraining the underlying string
theory. However, Supergravity (SUGRA)
constraints
\cite{sugra-constr_1,sugra-constr_2,sugra-constr_3}
imply the existence of severe
difficulties for the realisation of at least one class (of the two
studied classes) of brane inflationary models
\cite{gss} proposed in the context of IIB string
theory. Moreover, constraints are also imposed on the allowed
radiation emitted from the produced cosmic
superstrings
\cite{gss}.

The difficulties in realising a scalar field based inflationary model,
which on the one hand is consistent with particle physics and
CMB
measurements, while on
the other hand it satisfies minimal naturalness criteria, may be
overthrown if one changes the gravitational sector of the theory,
adding for instance an $R^2$-term
\cite{Starobinsky:1980te,Gottlober:1990um},
without changing its particle
content. This approach avoids on the one hand the constraints imposed
due to the generically formed cosmic
(super)strings,
which remain
undetectable \cite{Ade:2013xla},
and on
the other hand the (by hand) addition of an extra singlet to play the
r\^ole of the inflaton.
Note that it is
certainly reasonable to expect corrections to the low-energy effective
gravitational action, as for instance within noncommutative spectral
geometry
\cite{max-mairi},
moreover models with high
curvature terms seem to be favoured by the recent Planck
CMB
measurements
\cite{Ade:2013_infl}.

In what follows, we will briefly review the constraints on
supersymmetric hybrid inflation,
the consistency of brane inflation
models, as well as the
constraints on the allowed radiation
of cosmic superstrings.
We will
then highlight some open questions which deserve to be addressed given
the precision of current astrophysical and particle physics
measurements and the plethora of cosmological scenarios.

\section{Supersymmetric hybrid inflation}\label{susy-,hybrid-infl}
Models based upon $N=1$ SUSY
contain
complex scalar fields which often have flat directions in their
potential, offering candidates for inflationary models. Within this
framework, hybrid inflation
driven by
F- or D-terms represents the standard inflationary model, which
leads \cite{Jeannerot:2003qv} generically to cosmic
string
formation at the end of the
inflationary era.

F-term inflation
can be naturally
accommodated in the framework of GUTs when a GUT gauge group, G$_{\rm
GUT}$, is broken down to the SM gauge group, G$_{\rm SM}$, at an
energy scale $M_{\rm GUT}$ according to the scheme
\begin{equation}
{\rm G}_{\rm GUT} \stackrel{M_{\rm GUT}}{\hbox to 0.8cm
{\rightarrowfill}} {\rm H}_1 \xrightarrow{9}{M_{\rm
infl}}{1}{\Phi_+\Phi_-} {\rm H}_2 {\longrightarrow} {\rm G}_{\rm SM}~,
\end{equation}
where $\Phi_+, \Phi_-$ is a pair of GUT Higgs superfields in
non-trivial complex conjugate representations, which lower the rank of
the group by one unit when acquiring non-zero vacuum expectation
value. The inflationary phase takes place at the beginning of the
symmetry breaking ${\rm H}_1\stackrel{M_{\rm infl}}{\longrightarrow}
{\rm H}_2$.  The gauge symmetry is spontaneously broken by adding
F-terms to the superpotential. The Higgs
mechanism
leads
generically \cite{Jeannerot:2003qv}
to
Abrikosov-Nielsen-Olesen strings, called F-term
strings.

F-term inflation
is based on the
globally supersymmetric renormalisable superpotential
\begin{equation}\label{superpot}
W_{\rm infl}^{\rm F}=\kappa  S(\Phi_+\Phi_- - M^2)~,
\end{equation}
where $S$ is a GUT gauge singlet left handed superfield and $\kappa$, $M$
 are two constants ($M$ has dimensions of mass) which can be taken
 positive with field redefinition.
The scalar potential as a function
 of the scalar complex component of the respective chiral superfields
 $\Phi_\pm, S$ reads
\begin{equation}
\label{scalpot1}
V(\phi_+,\phi_-, S)= |F_{\Phi_+}|^2+|F_{\Phi_-}|^2+|F_ S|^2
+\frac{1}{2}\sum_a g_a^2 D_a^2~.
\end{equation}
The F-term is such that $F_{\Phi_i} \equiv |\partial W/\partial
\Phi_i|_{\theta=0}$, where we take the scalar component of the
superfields once we differentiate with respect to $\Phi_i=\Phi_\pm,
S$. The D-terms are $D_a=\bar{\phi}_i\,{(T_a)^i}_j\,\phi^j +\xi_a$,
with $a$ the label of the gauge group generators $T_a$, $g_a$ the
gauge coupling, and $\xi_a$ the Fayet-Iliopoulos (FI) term. By
definition, in the F-term inflation
the real constant $\xi_a$ is zero; it can only be non-zero if $T_a$
generates an extra U(1) group.  In the context of F-term hybrid
inflation the F-terms give rise to the inflationary potential energy
density while the D-terms are flat along the inflationary trajectory,
thus one may neglect them during inflation.

D-term inflation
has been studied a
lot in the literature, in particular since it can be naturally
implemented within SUSY GUTs, SUGRA, or string theories. The gauge
symmetry is spontaneously broken by introducing FI D-terms. In
standard D-term inflation, the constant FI
term
gets compensated by a
single complex scalar field at the end of the inflationary era,
leading to the formation of cosmic strings, called D-term
strings.

Standard D-term inflation
 is realised
within a scheme, such as
\begin{equation}
{\rm G}_{\rm GUT}\times {\rm U}(1) \stackrel{M_{\rm GUT}}{\hbox to
0.8cm{\rightarrowfill}} {\rm H} \times {\rm U}(1) \xrightarrow{9}{M_{\rm
nfl}}{1}{\Phi_+\Phi_-} {\rm H} \rightarrow {\rm G}_{\rm SM}~.
\end{equation}
It is based on the superpotential
\begin{equation}\label{superpoteninflaD}
W=\lambda S\Phi_+\Phi_-~,
\end{equation}
where $S, \Phi_+, \Phi_-$ are three chiral superfields and $\lambda$
is the superpotential coupling. It assumes an invariance under an
Abelian gauge group $U(1)_\xi$, under which the superfields $S,
\Phi_+, \Phi_-$ have charges $0$, $+1$ and $-1$, respectively. It
also assumes the existence of a constant FI term $\xi$.  

Note that a supersymmetric description of D-term inflation is
insufficient since the inflaton field reaches values of the order of
at least the Planck mass, even if one concentrates around only the
last 60 e-folds of inflation. Hence, the correct D-term inflation
analysis should be within the SUGRA
framework \cite{Rocher:2006nh}.
Moreover, D-term inflation must
be described with a non-singular formulation of supergravity when the
superpotential vanishes. To construct a formulation of SUGRA with
constant FI terms from superconformal theory, one finds \cite{toine1}
that under U(1) gauge transformations in the directions in which there
are constant FI terms $\xi_\alpha$, the superpotential $W$ must
transform as $\delta_\alpha W=\eta_{\alpha i}\partial^i W = -i
(g\xi_\alpha/M_{\rm Pl}^2)W$; one cannot keep any longer the same
charge assignments as in standard supergravity. D-term inflationary
models can be built with different choices of the K\"ahler
geometry \cite{Rocher:2006nh}.

Since cosmic strings \cite{Sakellariadou:2006qs}
 formed within a
large class of high energy physics models (SUSY, SUGRA and string
theories), one should consider mixed cosmological perturbation models
where the dominant r\^ole is played by the inflaton field and cosmic
strings have a small, but not negligible, contribution \cite{fpam}.
Restricting ourselves to the angular power spectrum we remain in the
linear regime, where \cite{fpam}
\begin{equation}
C_\ell =   \alpha     C^{\scriptscriptstyle{\rm I}}_\ell
         + (1-\alpha) C^{\scriptscriptstyle{\rm S}}_\ell~;
\label{cl}
\end{equation}
$C^{\scriptscriptstyle{\rm I}}_\ell$ and $C^{\scriptscriptstyle {\rm
S}}_\ell$ denote the (COBE normalised) Legendre coefficients due to
adiabatic inflaton fluctuations and those stemming from the cosmic
strings network, respectively and the coefficient $\alpha$ is a free
parameter giving the relative amplitude for the two contributions.
Comparing the $C_\ell$, given by Eq.~(\ref{cl}), with data obtained
from the CMB measurements, one imposes constraints on the string
tension \cite{fpam,str1,str2}.
The most recent Planck
collaboration data impose severe constraints on the string
contribution. More precisely, denoting by $f_{10}$ the fraction of the
spectrum contributed by cosmic strings
at a multipole of $\ell=10$, the Planck data imply
\cite{Ade:2013xla}
for the Nambu-Goto
string model that
\begin{equation}
f_{10}<0.015 \ \ {\rm leading\  to}\ \  G\mu/c^2 <1.5\times 10^{-7}~,
\end{equation}
with $G$ the Newton's constant and $\mu$ the string tension.\\ The
corresponding constraints for the Abelian-Higgs field theory model
read \cite{Ade:2013xla}
\begin{equation}
f_{10}<0.028 \ \ {\rm leading\  to}\ \  G\mu/c^2 <3.2\times 10^{-7}~.
\end{equation}

\section{Brane inflation}\label{brane-infl}
Within IIB string theory there are two classes of brane
inflation
models: D3/D7 inflation and
brane-antibrane inflation
with D3/${\overline{{\rm D}3}}$ being the most studied example. The
former is a string realisation of D-term
inflation,
where the r\^ole of a FI
term
is played by a
non-self-dual flux on the D7-brane. The two (D3 and D7) branes are
attracted to each other due to the breaking of
SUSY
by the FI term; inflation takes
place in an unwarped background.  The latter is a string realisation
of F-term inflation,
which is in
general plagued by the $\eta$-problem. The brane is attracted to the
antibrane due to warping. D3/${\overline{{\rm D}3}}$ inflation takes
place in a warped throat, as for instance the Klebanov-Strassler (KS)
geometry;
the warping
resolves the $\eta$-problem of the standard F-term inflation.

To accept the two classes of brane inflation
models studied within IIB string theory, one must firstly
examine their theoretical consistency. More precisely,
SUGRA
is a consistent theory, in the
sense that there is a globally well-defined Ferrara-Zumino multiplet,
provided (i) the FI terms
are
field-dependent and (ii) the moduli space are non-compact. Assuming
the SUGRA
 constraints remain valid in the
full string theory, we will examine the consistency of the two classes
of brane inflation and deduce \cite{gss1}
severe consequences for the dynamics of the inflationary mechanism in
the D3/D7 case.  Note that often in brane inflation, it is desirable
to leave the brane positions unfixed while the volume of
compactification is stabilised.  To examine whether the available
brane inflation models are consistent with the
SUGRA
constraints, one has to consider
the details of compactification and moduli stabilisation. Only once we
are convinced that the brane inflation models satisfy the
SUGRA
constraints, it makes sense, as a
second step, to investigate the observational consequences of the
models and test their observational consequences against current
measurements.

In the D3/D7 brane inflation model, a D3 brane is parallel to a D7
brane in the four noncompact directions, with the other legs of the D7
wrapping K3. The presence of a non-self-dual flux on the D7 brane
leads to SUSY
breaking and the
appearance of an attractive potential between the branes. The inflaton
field is given by $\phi=x^4+ix^5$ where $x^4$ and $x^5$ are the
directions perpendicular to both branes, along which they feel the
attractive potential. Inflation ends in a waterfall regime, when the
open string modes stretched between the branes become tachyonic.  The
FI term<
which is due to a
non-self-dual flux on the D7 brane, is field-dependent
\cite{gss1}
and given by
\begin{equation}
\xi={\delta_{\rm GS}\over {\rm vol}({\rm K}3)}~,
\end{equation}
where $\delta_{\rm GS}$ is the Green-Schwarz (GS) parameter. Let us
note that this is a rather standard result in string theory, where
field-dependent FI terms arise from GS anomaly cancellation. The
K\"ahler modulus $s={\rm vol(K3)}+iC_{(4)}$ (where $C(4)$ stands for
the 4-form) plays the r\^ole of the axion-dilaton, leading to a ${\rm
vol(K3)}$ dependence of $\xi$. At a first glance, this dependence
implies that the first SUGRA constraint is satisfied. However, for
this to be true the real part of the K\"ahler modulus, ${\rm
vol(K3)}$, must remain unfixed, or more specifically, the volume of K3
cannot be stabilised above the SUSY breaking scale
\cite{sugra-constr_3},
which is given by $\xi$
\cite{gss}:
\begin{equation}
V\sim g^2\xi^2=g^2{\delta^2_{\rm GS}\over [{\rm
vol(K3)}]^2}={\delta_{\rm GS}^2\over [{\rm vol(K3)}]^3}~.
\end{equation}
Hence, the scale of compactification $[{\rm vol(K3)}]^{-1/6}$ has to
be stabilised above the SUSY breaking scale which is approximately
$[{\rm vol(K3)}]^{-3}$.  The problem with D3/D7 brane inflation can be
also realised if one makes the remark that successful inflation
requires stabilisation of the K\"ahler modulus at a large finite value
above the SUSY
breaking scale, thus
rendering the FI term constant. In addition, this would imply that the
moduli space is compact.

One may thus conclude that the current scenario of D3/D7 brane
inflation is inconsistent with the SUGRA constraints. A possible way
out is to assume that quantum corrections may render the real part of
the K\"ahler modulus dependent not only on the ${\rm vol(K3)}$ but
also on the D3 position, or in other words on the brane separation
$r$. This implies that the FI term $\xi$ depends on $r$, which renders
the inflationary mechanism problematic since both the bifurcation
point and the Hubble constant during inflation depend on $\xi$.

In the D3/${\overline{{\rm D}3}}$ brane inflation model, the
brane-antibrane system lives at a specific point of a Calabi-Yau (CY)
manifold, and the net attractive force between the brane and antibrane
results from supersymmetry breaking. Here the brane-antibrane
separation plays the r\^ole of the inflaton. To accommodate a
suffiecient number of e-foldings, the first and second slow roll
parameters $\epsilon$ and $\eta$ must be very small and since $\eta$
cannot be much smaller than unity in a flat space, D3/${\overline{{\rm
D}3}}$ brane inflation is realised in a warped geometry; the
corresponding model is referred to as ${\mathbb K}$L${\mathbb M}$T
\cite{kklmmt}.
Since this inflationary model is a stringy realisation of F-term
inflation, there is no FI term, hence the first SUGRA constraint is
not applicable. The ${\mathbb K}$L${\mathbb M}$T D3/${\overline{{\rm
D}3}}$ brane inflation model can be made consistent with the second
SUGRA constraint, in the sense that the moduli space of the theory can
be made non-compact, through a SUSY breaking mechanism which in this
model can be set at a scale independent of the volume modulus. Hence,
the volume modulus can be fixed below the SUSY breaking scale,
something which cannot be achieved for D3/D7 brane inflation.

\section{Radiation of cosmic (super)strings}
Cosmic superstrings \cite{Sakellariadou:2008ie}
gravitational, axionic (Ramond Ramond (RR) or Neveu-Schwarz
Neveu-Schwarz (NS NS) particles, since cosmic superstrings are charged
under the two-forms $B_2$, $C_2$ which are Hodge dual to axionic
scalars in four dimensions), or dilatonic radiation.  As we have
already discussed, to render cosmic strings compatible with the CMB
data, the string tension gets severely constrained. In contrast to
their gauge theory analogues, cosmic superstrings have in general
Planck-scale energies and hence their high tension leads to
$G\mu/c^2\sim 1$ \cite{ew}
rendering them inconsistent with the CMB data . One
should thus conclude that string inflation models leading to the
production of cosmic superstrings with such high tensions must be
ruled out. Hence, the IIB string theory motivated inflationary models
that could be further examined are those realised within a warped
geometry, as for istance brane-antibrane inflation in a throat, or
models with large extra dimensions, since only in these cases the
cosmic superstring tension gets suppressed.

In what follows we will briefly summarise
\cite{gss1}
 the allowed channels of cosmic
superstring radiation in warped backgrounds given our previous
discussion (i) on the consistency of brane inflationary models with
respect to the SUGRA constraints and (ii) on the constraints on the
cosmic superstring tension.  Taking into account a consistent
compactification, we will see that severe constraints are imposed on
the allowed radiation emitted from cosmic superstrings. Let us note
that for gauge cosmic strings, gravitational radiation \cite{gw1,gw2}
is the main
channel of energy loss, allowing a cosmic string network to reach a
scaling solution.

Cosmic superstrings produced at the end of brane infation can be of
three types: F-strings (the fundamental strings of the theory),
D-strings (one-dimensional Dirichlet branes), or $(p,q)$ string bound
states of $p$ F-strings and $q$ D-strings. Considering a $(p,q)$
string in a KS throat, the gravitational power radiated per unit solid
angle is proportional to the square of the tension
$T_{(p,q)}=\sqrt{T_{\rm D}^2+T_{\rm F}^2}$, with $T_{\rm D}$ and
$T_{\rm F}$ denoting the tensions of the D-string and F-string,
respectively \cite{gss1}.
Further, since $(p,q)$ strings are charged under the $B_2^{\rm
NS}$ and $C_2^{\rm RR}$ 2-forms, they may emit massless RR or NS-NS
particles; the former are referred to as axions since the RR 2-form
$C_2$ is Hodge dual in four dimensions to the pseudo-scalar axion.
 
Considering axion emission from D-strings, it was shown
\cite{firouzjahi}
that the axionic
radiation is
\begin{equation}
P_{\rm RR}={\Gamma_{\rm RR}\mu_1^2\over \pi^2 g_{\rm s}\beta M_{\rm P}^2}~,
\end{equation}
while the gravitational one reads
\begin{equation}
P_{\rm g}=\Gamma_{\rm g} G\left({h^2\mu_1\over g_{\rm s}}\right)^2~,
\end{equation}
leading to the ratio
\begin{equation}
{P_{\rm RR}\over P_{\rm g}}=\left({8 \Gamma_{\rm RR}\over\pi
\Gamma_{\rm g}} \right){g_{\rm s}\over \beta h^4}~,
\end{equation}
where $\Gamma_{\rm RR}, \Gamma_{\rm g}$ are numerical factors of the
order ${\cal O}(50)$, $g_{\rm s}$ stands for the string coupling, $h$
denotes the warp factor and $\beta$ parametrises the difference in
normalisations between the Chern-Simmons and the Einstein-Hilber term
in the presence of warping. Thus, in the limit where $g_{\rm s}\ll 1$
and warping is negligible, power loss by gravitational radiation
dominates over axionic radiation. However, in a warped geometry, as in
the case of a KS background, since $h$ can be much smaller than unity,
one may naively conclude that axionic radiation becomes the dominant
one. We will return to this point later in our discussion.

Considering F-strings, the NS-NS particle emission reads
\begin{equation}
P_{\rm NS-NS}={\Gamma_{\rm NS}\mu_1^2 g_{\rm s}\over\pi^2\beta M_{\rm P}^2}~,
\end{equation}
which implies that the ratio of the axionic to the gravitational radiation is
\begin{equation}
{P_{\rm NS-NS}\over P_{\rm g}}=\left({8\Gamma_{\rm NS-NS}\over \pi
\Gamma_{\rm g}}\right){g_{\rm s}^3\over \beta h^4}~.
\end{equation}
Hence, the NS-NS particle emission is suppressed as compared to the RR
particle radiation.

The next interesting question is whether particle radiation can be
dominant over the gravitational one in the case of $(p,q)$
strings. This is an important issue since it may highlight an
observational difference between comic superstrings and their gauge
analogues, at least in the case that the former are produced in a
warped throat. Let us first remark that, as we have pointed out in
Ref.~\cite{gss1}.
the
enhancement of RR particle emission claimed for D-strings in
Ref.~\cite{firouzjahi}
is due to the
effect of warping. However, as discussed in
Ref.~\cite{gss1},
allowed radiation channels are constrained from the orientifold
projection imposed by a consistent flux compactification. The only
known consistent compactification of the Klebanov-Strassler geometry
is the flux compactification proposed by Giddings, Kachru and
Polchinski (GKP) in Ref.~\cite{gkp},
which
involves an orientifold projection ${\cal O}$ whose action on the
NS-NS and RR two-forms reads
\begin{equation}
{\cal O}B_2=-\sigma^\star B_2 \ \ \ \ {\rm and}\ \ \ \ {\cal
O}C_2=-\sigma^\star C_2~,
\end{equation}
respectively, where $\sigma^\star$ is the pull-back of an isometric
and holomorphic involution $\sigma$.  Noting that the internal
symmetry $\sigma$ acts on the internal manifold while it leaves the
four-dimensional non-compact space invariant, one concludes that the
NS-NS and RR two-forms with legs in the non-compact directions are
projected out \cite{projection}.
This
observation has important consequences for the allowed radiation
channels. More precisely, since the zero modes of $B_{\mu\nu}$ and
$C_{\mu\nu}$ do not appear in the spectrum
\cite{cmp},
there can be no massless RR or
NS-NS axions \cite{gss1}.
As
it has been argued in Ref.~\cite{gss1},
neither D- nor F-strings can lead to significant axionic emission,
since by construction of the brane inflationary model within which
they are formed, D- and F-strings will not have any massless axionic
radiation.  However, for a $(p,q)$ string arising from a wrapped
D3-brane, the situation may be different. In this case, while NS-NS
particle radiation is completely ruled out, RR radiation may be
possible since $(p,q)$ strings are also charged under the RR four-form
$C_4$.

However, to determine the allowed modes of radiation by cosmic
superstrings
within a warped
geometry,
one should consider not only
which modes will survive the orientifold projection, but one should
also keep in mind the correct dimensional reduction of those modes. As
it has been pointed out in
Ref.~\cite{Underwood:2010pm},
the
wave-function of the axionic zero mode is nontrivially modified by
warping, however there is only limited progress in understanding this
modification.  Since the resulting wave-function would affect the
magnitude of any radiation in this mode, lack of knowledge of the
modified wave-function implies that we are not yet in a position to
quantify the amplitude of the radiation.

Finally, cosmic superstrings may also emit dilatonic
radiation.
As is well-known,
massive dilatons are compatible with astrophysical observations only
if they acquire a Vacuum Expectation Value (VEV) before
nucleosynthesis.
Indeed, the
dilaton
gets a non-trivial potential in the
GKP compactification, however the mass of the dilaton located in the
throat will be suppressed by the warp factor.
  
In a warped geometry,
the
constraints \cite{Babichev:2005qd} 
on cosmic
superstring
tension as a function
of the dilaton mass are weakened \cite{Sabancilar}
since the mass gets suppressed by the
warp factor and the dilaton wave-function is localised in the throat
with an exponential fall off in the bulk, resulting to an enhancement
of the dilaton to matter coupling.

\section{Concluding remarks}
Cosmic (super)strings seem to be generically formed at the end of an
inflationary era in the context of inflationary models built within
supersymmetric grand unified theories, supergravitry, or even string
theory models. These objects decay emitting gravitational waves and
particles, the gravitational radiation being, most probably, the dominant decay
channel. 

As a contradiction to our theoretical expectations and despite various efforts we have not yet identified
any such objects in the sky, while temperature anisotropies data
constrain severely their potential cosmological r\^ole in structure
formation. As a result, the allowed string tension decreases
constantly rendering cosmic (super)strings weaker and therefore
diminishing any gravitational implications they may have.  However,
since the inflationary scale is related to the string tension, one
cannot freely constrain the tension of the strings without questioning
the implications for the inflationary model within which strings were
formed.

Conventional and well-studied inflationary models face some questions,
with the onset of inflation and the origin of the inflaton field being
the most important ones. To this list we have added one more, namely
the compatibility between the constraints on string tension and the
validity of the currently well-studied inflationary models. Absence of
topological defects may imply lack of knowledge of the thermal history
of the universe, which will certainly affect our understanding of
inflation.

Let us end by stressing that our conclusion is not to rule out
inflation but to emphasise that a number of the currently available
models may turn out to be unphysical or inconsistent with the theoretical
framework upon which they were based.

\section*{References}
\bibliographystyle{jphysicsB}

\bibliographystyle{jfm2}


\end{document}